\documentclass[%
superscriptaddress,
longbibliography,
 amsmath,
 amssymb,
 aps, 
 twocolumn,
]{revtex4-1}

\usepackage{graphicx}
\usepackage{dcolumn}
\usepackage{xr}
\usepackage{bm}
\usepackage[colorlinks=true,
  allcolors={blue}]{hyperref}
  \usepackage{multirow}
  \usepackage{braket}
\usepackage{iftex}

\usepackage{xcolor}

\newcommand{\fgt}{Fe$_3$GeTe$_2$}

\usepackage{changes}

\usepackage[resetlabels,labeled]{multibib}

\begin{document}


\title{Phonon-induced renormalization of exchange interactions in metallic two-dimensional magnets}

\author{Danis I. Badrtdinov}
\affiliation{Institute for Molecules and Materials, Radboud University, Heijendaalseweg 135, NL-6525 AJ Nijmegen, The Netherlands}

\author{Mikhail I. Katsnelson}
\affiliation{Institute for Molecules and Materials, Radboud University, Heijendaalseweg 135, NL-6525 AJ Nijmegen, The Netherlands}

\author{Alexander N. Rudenko}
\email{a.rudenko@science.ru.nl}
\affiliation{Institute for Molecules and Materials, Radboud University, Heijendaalseweg 135, NL-6525 AJ Nijmegen, The Netherlands}

\date{\today}

\begin{abstract} 
The presence of spin-polarized charge carriers in metallic magnets provides a mechanism for spin-lattice interactions mediated by electron-phonon coupling. Here, we present a theory of this mechanism used to estimate its effect on the exchange interactions in 2D magnets. Starting from a square lattice model at half filling, we show that the presence of electron-phonon coupling with equilibrium phonon distribution leads to a notable suppression of exchange interactions with temperature.
We then apply our approach to the prototypical 2D metallic ferromagnet, \fgt{}, with moderate electron-phonon coupling. We find that the exchange interactions undergo a renormalization, leading to a softening of the magnon modes, and suppression of the Curie temperature by $\sim$10\%. We expect that this effect can be further enhanced in systems with strong electron-phonon coupling, as well as for non-equilibrium distribution of phonons induced by strong laser fields or charge currents.

\end{abstract}

\maketitle

{\it Introduction.}---  Magnetism is a quantum phenomenon that plays an important role in technology. Although fundamental aspects of magnetism had been established decades ago,
understanding the behavior of magnetic materials at the microscopic scale is still the subject of intensive research. Recently, the interest to fundamental
aspects of magnetism has been revived due to the emergence of novel materials that host low-dimensional forms of magnetism, allowing for its efficient manipulation
and control \cite{Gong2017,Huang2017,Klein2018,Huang2018,Jiang2018}. Among these materials, metallic two-dimensional (2D) magnets with the most prominent example of \fgt{} \cite{Deng2018,Fei2018} constitutes a special class of objects where charge carriers provide additional degree of freedom for tailoring the magnetic properties.

One of the most basic models in magnetism is the Heisenberg model 
\begin{equation}
{\cal H}_0 = \sum_{i>j}J_{ij} \mathbf{S}_i \mathbf{S}_j  
\label{eq:Heisenberg_hamiltonian}
\end{equation}
that describes interaction between magnetic moments ${\bf S}_i$ via the pairwise
exchange interaction $J_{ij}$ determined by the electronic structure of a material.
A standard way to calculate $J_{ij}$ is to employ the so-called magnetic force theorem, yielding a convenient expression in the form \cite{liechtenstein1987,PhysRevB.61.8906,Szilva2023}
\begin{equation}
    J_{ij} = 2 \, \mathrm{Tr}_{\omega L} \left[  \Delta_i G_{ij}^{\uparrow} (i\omega_n) \Delta_j G_{ji}^{\downarrow} (i\omega_n) \right] S^{-2},
    \label{eq:Jij}
\end{equation}
where $G_{ij}^{\sigma}(i\omega_n)$ is the spin-polarized electron propagator, $\Delta_i$ is the exchange splitting of electrons at lattice site $i$, and $\mathrm{Tr}_{\omega L}$ denote the sum over Matsubara frequencies $i\omega_n$ and orbital indices $L$.
In most of the practical situations, $\Delta_i$ and $G_{ij}^{\sigma}$ are derived from single-particle Hamiltonians, disregarding explicit treatment of many-body effects.
A modification of Eq.~(\ref{eq:Jij}) has been proposed in the context of strongly correlated systems \cite{PhysRevB.61.8906}, giving the possibility to take into account renormalization of $J_{ij}$ due to  electron-electron interactions. In this case $\Delta$ is replaced by spin-dependent part of the local electron self-energy $\Sigma^{\sigma}(i\omega_n)$, obtained self-consistently within  dynamical mean-field theory (DMFT). This self-energy accounts for dynamical spin fluctuations and ensures temperature-dependence of the exchange.

An alternative way to describe temperature dependence of $J_{ij}$ due to magnetic fluctuations is to introduce magnetic disorder within the formalism of disordered local moments \cite{Pindor_1983,Staunton_1985,Staunton_1986}. The degree of disorder is temperature dependent and typically determined from atomistic spin simulations for each temperature. Subsequently, Eq.~(\ref{eq:Jij}) is modified to take the disorder into account \cite{Henk2012,Szilva2013}. 
This approach can be considered as a semiclassical approximation to DMFT 
\cite{PhysRevB.67.235105}.
Recently, semianalytical methods have been proposed to treat this problem, in which temperature-dependent corrections to the exchange are evaluated from spin correlation functions 
\cite{Rozsa2023}.

At nonzero temperatures, thermal lattice fluctuations provide additional contribution to the temperature dependence of exchange interactions \cite{Ebert2020}. 
This effect is supposed to be more pronounced in low dimensions where electronic structure is more susceptible to perturbations of the atomic structure. Earlier attempts to describe spin-lattice
effects in 2D magnets were limited to insulators, such as CrI$_3$ \cite{Delin2022},
with the main effect originating from a direct modification of the hopping integrals upon atomic displacements. In contrast, metallic magnets allow for a different mechanism of spin-lattice interaction related to the coupling of phonons with conducting electrons that is strongly temperature dependent.  

In this Letter, we study temperature-dependent renormalization of exchange interactions in 2D materials induced by the electron-phonon coupling. We propose to generalize Eq.~(\ref{eq:Jij}) for the case of electron-phonon coupling. To this end, we perform renormalization of the Green's functions using the Dyson equation
$G_{\mathbf{k}}^{-1}(i \omega_n) \rightarrow \widetilde{G}_{\mathbf{k}}^{-1}(i \omega_n) = G_{\mathbf{k}}^{-1}(i \omega_n) - \Sigma_{\mathbf{k}}(i \omega_n)$, which also leads to renormalization of the exchange splitting $ \Delta \rightarrow \widetilde{\Delta}_{\mathbf{k}}(i \omega_n) = \Delta + \Sigma_{\mathbf{k}}^{\uparrow}(i \omega_n) - \Sigma_{\mathbf{k}}^{\downarrow} (i \omega_n)$,
where we introduced the spin-polarized self-energy to be evaluated via the standard expression \cite{Migdal1958,AGD_book,mahan_book}
\begin{equation}
    \Sigma^{\sigma}_{\mathbf{k}} (i \omega_n) = -T \sum_{\mathbf{k}^\prime \nu m} G^{\sigma}_{\mathbf{k}}(i \omega_n) \vert g^{\nu \sigma}_{\mathbf{k} \, \mathbf{k}^\prime}\vert ^2 D_{\mathbf{k - k^\prime}}(i \omega_n - i \omega_m),
    \label{eq:Migdal_approximation}
\end{equation}
which connects the electron $G_{\mathbf{k}}(i \omega_n)$  and phonon $D^{\nu}_\mathbf{q}(i \omega_n)$ propagators defined on the fermionic Matsubara frequency axis $\omega_n = \pi T(2n+1)$ via the interaction vertex $g^{\nu \sigma}_{\mathbf{k} \, \mathbf{k}^\prime}$ at temperature $T$. The vertex corrections can be neglected according to the Migdal's theorem \cite{Migdal1958}, due to smallness of phonon frequencies in comparison with typical electron energies. 

\begin{figure}[tbp]
\centering
\includegraphics[width=1.0\linewidth]{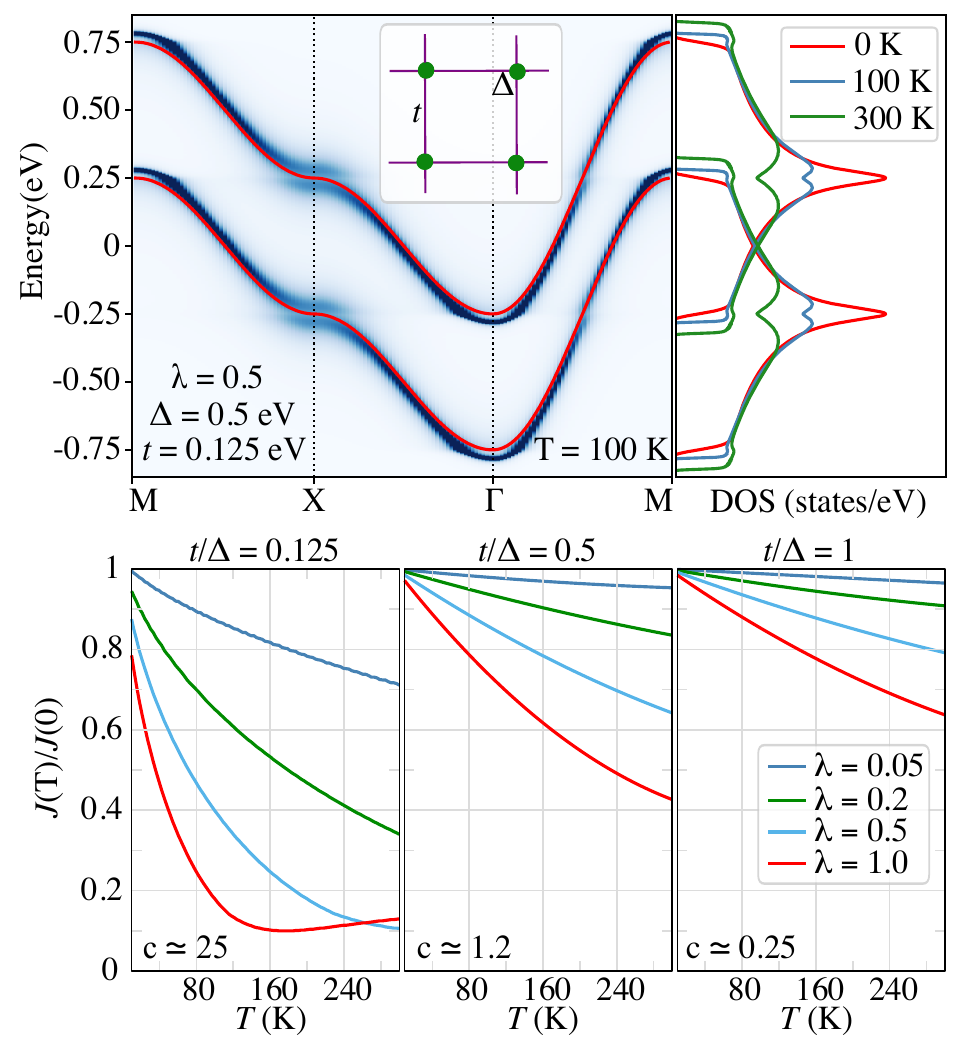}
\caption{(Top) Electron spectral functions and densities of states (DOS) for a square lattice calculated in the presence of the electron-phonon interactions at different temperatures. (Bottom) Temperature dependence of exchange interaction renormalization in square lattice  calculated for different coupling constants $\lambda$ and $t/ \Delta$ ratios. $c$ is the renormalization constant introduced in Eq.~(\ref{eq:JT}).
}
\label{fig:Square_lattice}
\end{figure}

{\it Square lattice}.--- To illustrate the effect of electron-phonon coupling on the magnetic exchange in metallic 2D magnets, we begin with a minimal model. Specifically, we consider spin-polarized electrons in a square lattice at half-filling interacting with acoustic phonons in the equilibrium. The model is described by the Hamiltonian,
\begin{eqnarray}
\notag
    H = t\sum_{\langle ij \rangle \sigma} c^\dagger_{i \sigma} c_{j \sigma} + \frac{\Delta}{2} \sum_{i} ( n^\uparrow_i - n^\downarrow_i) + \\
   + \sum_{\bf q} \omega_{\bf q} b^{\dag}_{\bf q} b_{\bf q} + \sum_{{\bf q},\langle ij\rangle\sigma} g_{\bf q} (b_{\bf q}^{\dag} + b_{-{\bf q}})c_{i\sigma}^{\dag} c_{j\sigma},
   \label{eq:tb_hamilt}
\end{eqnarray}
where $c_{i \sigma}^\dagger (c_{i\sigma})$ and $b_{\bf q}^\dagger (b_{\bf q})$ are the creation (annihilation) operator of electrons at site $i$ and spin $\sigma$, and phonons with wave vector ${\bf q}$ and frequency $\omega_{\bf q}$, $n_i^\sigma = c_{i \sigma}^\dagger c_{i\sigma}$, $t$ is the nearest-neighbor hopping, and $\langle ij \rangle$ labels the nearest-neighbour pairs. $\Delta$ is the on-site interaction giving rise to the exchange splitting of electronic bands, and $g_{\bf q}$ is the electron-phonon coupling. In the insulating state determined by the condition $t/\Delta < 0.125$, Eq.~(\ref{eq:Jij})
results in $J \simeq 4t^2/\Delta$ according to the Anderson superexchange theory~\cite{Anderson1959}. In the metallic state ($t/\Delta \ge 0.125$)
the electron-phonon coupling is, in general, not negligible, and must be taken into account.

For simplicity, we approximate phonon dispersion by a simple Debye-like model $\omega_{\bf q} = v q$ keeping in mind $\hbar \omega_D \gg k_B T$, where $\omega_D = v \pi/a$ is the characteristic frequency and $v$ is the sound velocity. The electron-phonon interaction matrix element is then takes the form $g_\mathbf{q} = g_0 \sqrt{\hbar q/(2M v})$, where $g_0$ is the interaction constant and $M$ is the nuclear mass. 
At long wavelengths and not too low temperatures ($\hbar v q\ll k_B T$) phonons can be considered classically with the occupation $\langle b^{\dag}_{\mathbf{q}}b_{\mathbf{q}}\rangle \simeq k_B T/\hbar v q$. Then the electron self-energy [Eq.~(\ref{eq:Migdal_approximation})] can be recast in a simple form as

\begin{equation}
\Sigma^\sigma_{\mathbf{k}}(\omega, T)  \simeq 2 \lambda \frac{k_B T}{N^\sigma_F}    \sum_{\mathbf{q}} G^\sigma_{\mathbf{k + q}}(\omega),
\label{eq:Selfen_phonon_highT}
\end{equation}
where $\lambda = g_{\bf q}^2 N^\sigma_F/\omega_{\mathbf{q}} = g_0^2 N^\sigma_F/2M v^2$ is the dimensionless electron-phonon coupling constant, $N^\sigma_F=\sum_{\bf k} \delta(\varepsilon_{\bf k}^{\sigma})$ is the electron densities of states (DOS) at the Fermi energy.  One can see that in this minimal model, temperature correction to the exchange is only determined by the constant $\lambda$ and by details of the electronic structure. 

In the top panel of Fig. \ref{fig:Square_lattice}, we show the spectral function $A^\sigma_{\mathbf{k}} (\omega, T) = -1/\mathrm{\pi} \, \mathrm{Im} [\widetilde{G}^\sigma_{\mathbf{k}}(\omega, T)]$ calculated for spin-polarized electrons in a square lattice in the presence of a moderate ($\lambda=0.5$) electron-phonon coupling at different temperatures. The coupling results in a ${\bf k}$-dependent band broadening and energy shifts of the band edges, which scale linearly with temperature.   The bottom panel of Fig. \ref{fig:Square_lattice} shows relative renormalization of the exchange interaction $J(T)/J(0)$ for different coupling constants $\lambda$ and $t/\Delta$ ratios. 
Although the spectral function temperatures does not display any qualitatively new features at finite temperature, even weak electron-phonon coupling results in a noticeable renormalization of $J_{ij}$. Under the assumptions made above, one can expand the many-body correction to the exchange keeping the linear terms only, yielding 
\begin{equation}
    J(T) = J(0) - c\lambda T,
    \label{eq:JT}
\end{equation}
where $c$ is a renormalization constant determined by the electronic structure of the system. This expression assumes that the electron-phonon contribution to the self-energy can be considered as a perturbation and taken into account in the linear approximation which seems to be reasonable in the most of cases. Importantly, the temperature correction $\delta J=-c \lambda T$ does not scale with $J(0)$ meaning that the renormalization effect is more significant when $J(0)\lesssim c\lambda T$.
As we are mostly interested in the ordered phase below the critical temperature, one can recast the renormalization criterion as $c \gtrsim T_\mathrm{c}/\lambda T$, keeping in mind the estimate $T_\mathrm{c}\sim J(0)$. In Fig. \ref{fig:Square_lattice}, we provide these constants fitted for the three cases considered.
We emphasize that the renormalization effect persists even for spin-independent electron-phonon coupling, i.e., when $\lambda^{\uparrow}=\lambda^{\downarrow}$. For unequal $\lambda^{\uparrow}$ and $\lambda^{\downarrow}$ the effect is expected to be stronger.

As one can see from Fig. \ref{fig:Square_lattice}, at the point of metal-insulator transition $t/\Delta$ = 0.125, even small $\lambda$ = 0.05 already leads to a $\sim$10 \% suppression of the exchange interaction at $T=50$ K, which increases further at higher temperatures reaching up to 90 \% for $\lambda \gtrsim 0.5$. At larger $t/\Delta$, the absolute value of $J_{ij}$ increases, which limits its renormalization due to electron-phonon coupling. In this case the effect can still be significant for high enough coupling constants and temperatures, and will be further enhanced provided $\lambda^{\uparrow} \neq \lambda^{\downarrow}$. Potentially strong temperature-dependence of exchange interactions suggests that it must be taken into account while calculating thermodynamic averages. In particular, we expect suppression of the ordering temperatures for magnetic systems with strong electron-phonon coupling.

\begin{figure}[tbp]
\centering
\includegraphics[width=1\linewidth]{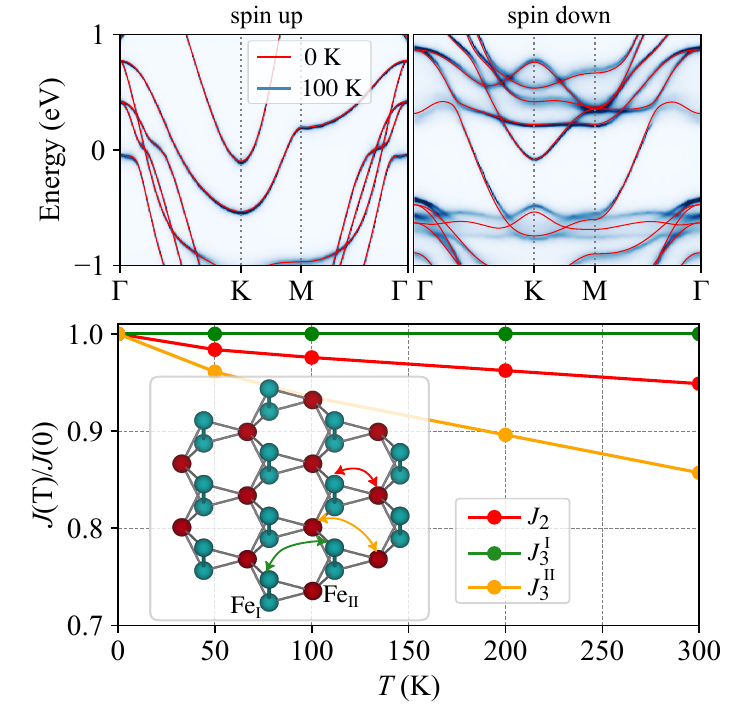}
\caption{(Top) Spin-resolved electron spectral function of monolayer \fgt{} calculated in the presence of the electron-phonon coupling at $T = 100$ K for the states near the Fermi level. Left and right panels correspond to the spin-up and spin-down states, respectively. Original single-particle DFT band structure ($T=0$) is shown by the red solid line. (Bottom) Temperature dependence of the renormalized leading exchange interactions shown schematically in the inset. The numerical values of $J_{ij}$ are given in the Supplemental Material~\cite{SM} (see also references~\cite{liechtenstein1987, mazurenko2005, Tyablikov_book, Nolting_book, Frobrich2006} therein).}
\label{fig:J_T}
\end{figure}

{\it Exchange renormalization in } \fgt{}.--- 
{One of well-known examples among 2D magnets is metallic \fgt{}, for which experimental studies demonstrate fingerprints of strong spin-lattice coupling~\cite{Magnon_phonon_Luojun2019, Mokrousov2023}.} Let us now estimate the role of electron-phonon coupling in the magnetic properties of this system using the approach presented above.  The magnetic lattice of \fgt{} is schematically shown in the inset of Fig. \ref{fig:J_T}, which exhibits two distinct positions Fe$_{\rm I}$ and Fe$_{\rm II}$ with magnetic moments 2.4 $\mu_B$ and 1.6 $\mu_B$~\cite{Pushkarev2023}. As a starting point, we calculate the electronic structure of \fgt{} within density functional theory (DFT) using the plane-wave pseudopotential method as implemented in {\sc Quantum Espresso ({\sc qe})}~\cite{QE1,QE2} package adopting the Perdew-Burke-Ernzerhof (PBE) exchange-correlation functional~\cite{PBE}. The phonon-related properties spectra are calculated within the density functional perturbation theory in the linear response regime~\cite{Baroni2001}.  In the calculations, we use the same numerical parameters as in our previous work~\cite{Badrtdinov2023} summarized in the SM \cite{SM}. We restrict ourselves to the monolayer structure with the experimental lattice parameters of bulk \fgt{}~\cite{Deiseroth2006}.    

The electronic structure of \fgt{} in the vicinity of the Fermi energy is mainly formed by Fe $(3d)$  hybridized with Te $(5p)$ states~\cite{Badrtdinov2023}, which allows us to  parameterize them by maximally localized Wannier functions~\cite{marzari1997, Marzari2012}. Using such parametrization, we construct a tight-binding Hamiltonian in the basis of Wannier functions to calculate the electron propagators as well as electron-phonon matrix elements needed to obtain the electron self-energy in the Migdal approximation (see the SM for details \cite{SM}).  For this purpose, we use the {\sc epw} code~\cite{Ponce2016, Giustino2017} modified to treat spin-polarized states. 
Fig. \ref{fig:J_T} shows spin-resolved spectral function calculated for monolayer \fgt{} at $T=0$ and $T=100$~K. One can see that spin-down electrons are more susceptible to the electron-phonon coupling, which is consistent with distinct coupling constants $\lambda^\uparrow \sim $ 0.5 and $\lambda^\downarrow \sim $ 0.26~\cite{Badrtdinov2023}.
Such behavior indicates a phonon-induced renormalization of the exchange splitting $\Delta_{\bf k}(\omega)$, which provides a direct contribution to the exchange interactions $J_{ij}$ according to Eq.~(\ref{eq:Jij}).

The resulting temperature dependence of the renormalized leading exchange interactions between iron atoms in \fgt{} is presented in Fig. \ref{fig:J_T} (see the SM for further details \cite{SM}). Overall, the effect of exchange renormalization is not large, and found to be within 10\% for $T=200$~K. This can be ascribed to a relatively large effective $\tilde{t} / \tilde{\Delta}$ ratio, in accordance with the results presented in Fig. \ref{fig:Square_lattice}.  Indeed, our first-principles calculations yield the average exchange splitting $\tilde{\Delta}$ of Fe($d$) electrons to be $\sim$2 eV, while the effective hopping $\tilde{t}$ determined from the bandwidth $W_d \simeq$ 10 eV is estimated to be $\tilde{t} \simeq W_d/2N_{nn} \approx$ 1.6 eV, where $N_{nn}=3$ is the number of nearest neighbors in the honeycomb lattice. The resulting effective parameters ratio $\tilde{t}/ \bar{\Delta} \sim $ 0.8 suggests a small renormalization effect, in agreement with Fig. \ref{fig:Square_lattice}. The corresponding renormalization constant $c$ introduced in Eq.~(\ref{eq:JT}) is estimated to be around 0.1--0.2 in \fgt{}, which also indicates weak renormalization. Despite the effect is relatively small, the
renormalization can still affect the observables such as the spectrum of spin-wave excitations or the Curie temperature. We note that Eq.~(\ref{eq:Jij}) for the exchange is only applicable to the ordered phase ($T<T_\mathrm{C}$), since in the paramagnetic phase both $\Delta_i$ and $\braket{S_i^z}$ tends to zero, and this limit requires a special careful treatment \cite{Pourovskii2016,Katanin2023}. Here we restrict ourselves only to the ferromagnetic phase leaving the problem of electron-phonon interaction in the paramagnetic phase for the future studies.

\begin{figure}[tbp]
\centering
\includegraphics[width=1\linewidth]{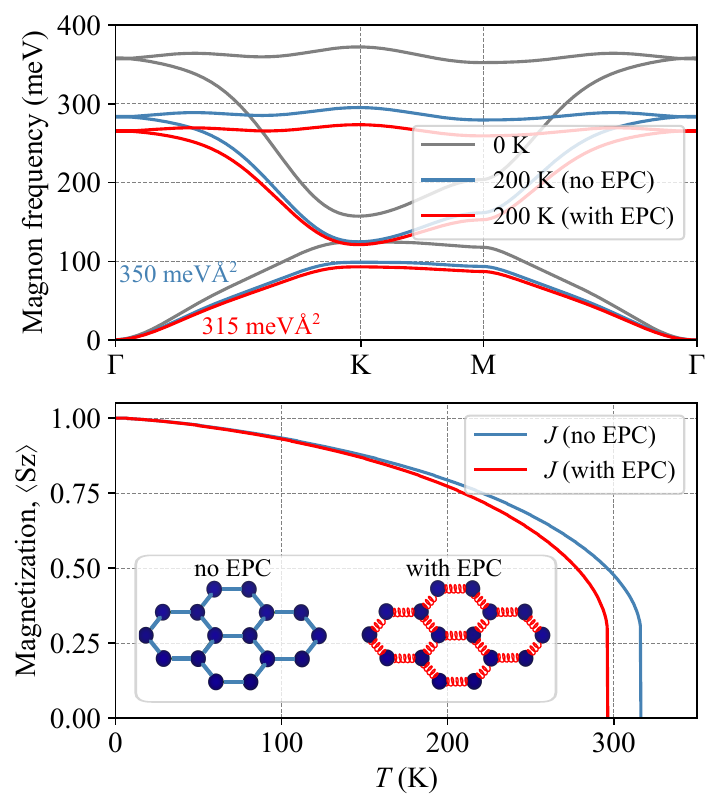}
\caption{(Top) Spin-wave spectra of \fgt{} calculated for different temperatures taking into account bare (no EPC) and renormalized (with EPC) exchange interactions $J$. The values in the left corner correspond to the spin stiffness. (Bottom) Temperature-dependent magnetization in \fgt{} calculated with and without renormalization by the electron-phonon coupling.} 
\label{fig:Magn_T}
\end{figure}

{\it Spin-wave renormalization in }\fgt{}.--- 
Let us now discuss the role of exchange renormalization in the stability of magnetic order.
For 2D magnets with long-range order, it is crucial to take magnetic anistropy into account, in accordance with the Mermin-Wagner theorem \cite{Mermin1966}. Here, we consider a spin-1 Heisenberg model in the presence of single-ion anisotropy (SIA) described by the Hamiltonian
\begin{equation}
{\cal H} = {\cal H}_0 + A \sum_{i} (S^z_{i})^2,
\end{equation}
where ${\cal H}_0$ is the isotropic term given by Eq.~(\ref{eq:Heisenberg_hamiltonian}), and $A \simeq -0.35$ meV is the anisotropy parameter~\cite{Badrtdinov2023}.

For a system with multiple magnetic sublattices, the magnons frequencies $\Omega_{{\bf q} \nu}$ can be calculated by diagonalization of the spin-wave Hamiltonian~\cite{Rusz2005},
\begin{equation}
{\mathcal{H}}^{SW}_{\mu \nu}(\mathbf{q})  = \left[ \delta_{\mu \nu} \left( 2A \Phi +  \sum_{\chi} J_{\mu \chi}(\mathbf{0}) \right) - J_{\mu \nu}(\mathbf{q}) \right] \braket{S_z},
\label{eq:SW}
\end{equation}
where $J_{\mu \nu}({\bf q})$ are the Fourier transform of the exchange interaction matrix, $\Phi = 1 - \left[ 1 - \braket{S^2_z}/2  \right]$ is the Anderson-Callen decoupling factor \cite{AndersonCallen,Frobrich2000} for $S=1$, which satisfies the kinematic condition. The indices $\mu$ and $\nu$ run from 1 to 3 over magnetic sublattices.  From Eq.~(\ref{eq:SW}) follows that the magnon energies $\Omega_{\bf q} \sim \braket{S_z}$, which is not always correct because the magnon excitations are known to exist beyond the critical temperature, assuming strong short-range magnetic order. Here, we are not interested in the paramagnetic regime, but note that the behavior at $T>T_\mathrm{C}$ can be restored by introducing short-range spin-order parameter as it is done, e.g., in  self-consistent spin-wave theory \cite{Irkhin1999}. 

Fig. \ref{fig:Magn_T} shows the calculated spin-wave spectra of monolayer \fgt{} that exhibit two optical branches and one acoustic branch dispersing quadratically $\Omega_{\bf q} \simeq \Omega_0 + D q^2$ near the $\Gamma$ point, where $D$ is the spin-stiffness constant and $\Omega_0$ is the SIA-induced gap of the order of meV.
Temperature dependence of the magnetization  $\braket{S_z}$ already leads to a softening of the magnon branches, which drops to 0 at $T = T_\mathrm{C}$.  As the same time, taking into account the temperature dependence of magnetic interactions increases this softening even stronger, reducing the spin-stiffness constant $D$.

At nonzero temperatures, the magnetization $\braket{S_z}$ is determined by spin-wave excitations, which suppress long-range magnetic order. Within the Green's function formalism, $\braket{S_z}$ can be determined from the equations \cite{Tyablikov_book,Nolting_book,Frobrich2006},
\begin{equation}
\langle (S_i^z)^n S_i^{-} S_i^{+} \rangle = \langle [S_i^+, (S^z_i)^nS_i^-] \rangle \sum_{{\bf q}\nu} \langle b_{{\bf q}\nu}^{\dag} b_{{\bf q}\nu} \rangle
\label{eq:Tyablikov}
\end{equation}
obtained by Tyablikov (also known as the RPA result). Here, $S^{-}_i$ and $S^{+}_i$ are the ladder operators, $\langle b_{{\bf q}\nu}^{\dag} b_{{\bf q}\nu} \rangle = [\mathrm{exp}(\Omega_{\bf q\nu}/k_BT)-1]^{-1} $ is the equilibrium magnon distribution, and $n=0,1$ for $S=1$. Equations (\ref{eq:Tyablikov}) and (\ref{eq:SW}) are to be solved self-consistently with respect to $\braket{S_z}$ and $\braket{S_z^2}$ (see the SM \cite{SM} for details).

The exchange parameters calculated at zero temperature lead to $T_\mathrm{C} \simeq$ 315 K, which overestimate the experimental temperature $T_\mathrm{C} \sim$ 200 K~\cite{Deng2018, Fei2018, Verchenko2015, Bin2013, Roemer2020}.  At the same time, considering the temperature-dependent renormalization of magnetic interactions, we obtain a $\sim$10\% reduction of the Curie temperature down to $T_\mathrm{C} \simeq$ 295 K. Although the resulting effect is not large, we observe a correct trend toward the experimental values. Other possible reasons for the discrepancy can be related to strong correlation effects addressed earlier in Ref.~\onlinecite{Ghosh2023}. 

{\it Discussion. ---}
It is natural to ask whether the renormalization of exchange could be enhanced in real materials.
For most of the systems, electron-phonon coupling can be treated adiabatically, reflecting the fact that typical phonon energies $E_{ph}$ are much smaller than typical electron energies $E_{el}$. In this situation, electron-phonon contribution to the self-energy is important only within a narrow energy layer of the order of $E_{ph}$ near the Fermi energy.
Hence, its contribution to the quantities integrated over the whole energy range has a smallness in the parameter $E_{ph}/E_{el}$ \cite{Brovman1974}. Therefore, one possible way to achieve stronger renormalization effect is to focus on systems for which the adiabatic approximation does not hold, i.e., when $E_{el} \lesssim E_{ph}$. Such systems may include, for instance, materials with narrow (or flat) electron bands close to the Fermi energy, which potentially leads to exceptionally strong electron-phonon coupling \cite{Hewson1981,KT1985, Lugovskoi2019}. However, theoretical description of this regime goes beyond the Migdal theorem Eq.~(\ref{eq:Migdal_approximation}), and requires a separate consideration. In this context, theory of the antiadiabatic limit developed for conventional superconductivity appears as a promising direction for further studies of the exchange renormalization \cite{Sadovskii2020}. 

Another possible approach to enhance the effect is to make use of out-of-equilibrium distributions in describing the coupled dynamics of electrons and phonons. Such distributions could be achieved by charge or heat currents flowing through the system, as well as by optically-driven excitations, leading to enhancement of the electron-phonon coupling \cite{Maldonado2017,Maldonado2020}. In the presence of strong visible or ultraviolet laser fields, corrections are required to compute the exchange interactions, whose contribution might lead to novel effects \cite{Mikhaylovskiy2015,secchi2013non,secchi2016non}. The role of adiabaticity in these cases is much less clear. One can also expect strong effects under infrared laser fields, keeping in mind the possibility to reach much larger atomic displacements than the equilibrium ones \cite{cavalleri2,cavalleri1}. 
Last but not the least, in this study we limited ourselves to the isotropic exchange interaction, which is typically the dominant magnetic interaction in real magnetic materials. Given that temperature corrections to the exchange obtained in this work do not scale with the exchange itself, one can expect stronger renormalization effects for anisotropic magnetic interactions, such as DMI. This case constitutes another interesting problem, which we leave for further studies. 

The work was supported by the European Union’s Horizon 2020 research and innovation program under European Research Council synergy grant 854843 “FASTCORR”.


%

\pagebreak
\widetext
\begin{center}
\textbf{\large Supplemental Materials: Phonon-induced renormalization of exchange interactions in metallic two-dimensional magnets}
\end{center}
\setcounter{equation}{0}
\setcounter{figure}{0}
\setcounter{table}{0}
\setcounter{page}{1}
\makeatletter
\renewcommand{\theequation}{S\arabic{equation}}
\renewcommand{\thefigure}{S\arabic{figure}}
\renewcommand{\bibnumfmt}[1]{[S#1]}
\renewcommand{\citenumfont}[1]{S#1}

Temperature-dependent exchange interactions were calculated using local force theorem approach~\cite{S_Ref1, S_Ref2}:
\begin{equation}
J_{ij}(T) =  -\frac{1}{2 \pi S^2}  \int \limits_{-\infty}^{E_F} d \omega  \,{\rm Im} \left[  \widetilde{\Delta}_i (\omega, T) \widetilde{G}_{ij}^{\downarrow} (\omega, T) \widetilde{\Delta}_j (\omega, T) \widetilde{G}_{ji} ^{\uparrow} (\omega, T) \right].
\label{eq:Exchange}
\end{equation}

Here $\widetilde{\Delta}_i (\omega, T)$ and $\widetilde{G}^{\sigma}_{ij} (\omega, T)$ stands for Fourier transformed component of correlated intra-orbital spin-splitting energy and Green's function due to electron-phonon coupling:
\begin{equation}
 \begin{aligned}
& \widetilde{\Delta}_i (\omega, \mathbf{k}, T) = H_{i}^{\uparrow}(\mathbf{k}) - H_{i}^{\downarrow}(\mathbf{k}) + \Sigma^\uparrow(\omega, \mathbf{k}, T)  - \Sigma^\downarrow (\omega, \mathbf{k}, T),  \\
& \widetilde{G}_{ij}^{-1} (\omega,  \mathbf{k}, T) = G_{ij}^{-1} (\omega,  \mathbf{k}) - \Sigma(\omega, \mathbf{k}, T).
\label{eq:Correlations}
\end{aligned}
\end{equation}

The electron self-energy of electron-phonon interaction in Migdal approximation has the form
\begin{equation}
 \Sigma^\sigma_{n \mathbf{k}}(\omega, T)  =  \sum_{m \,\mathbf{q} \nu } |g^\sigma_{mn, \nu}(\mathbf{k}, \mathbf{q})|^2  \times   \left[ \frac{b_{\mathbf{q}\nu} + f^\sigma_{m  \mathbf{k + q}}}{\omega - \varepsilon^\sigma_{m  \mathbf{k + q}} + \hbar  \omega_{\mathbf{q \nu}} + i\eta}  + \frac{b_{\mathbf{q}\nu} +1 - f^\sigma_{m  \mathbf{k + q}}}{\omega - \varepsilon^\sigma_{m  \mathbf{k + q}} - \hbar  \omega_{\mathbf{q \nu}} + i\eta} \right]  ,  
\label{eq:Selfen_phonon}
\end{equation}
where $b_{\mathbf{q}\nu} = (\exp[\hbar \omega_{\mathbf{q \nu}}/k_BT] - 1)^{-1}$ corresponds to the Bose occupation function for phonons with wave vector $\mathbf{q}$, mode index $\nu$ and frequency $\omega_{\mathbf{q \nu}}$. In turn, $f^\sigma_{m  \mathbf{k}} = (\exp[(\varepsilon^\sigma_{m  \mathbf{k}})/k_BT] + 1)^{-1}$ is the Fermi occupation function for the electron states with energy $\varepsilon^\sigma_{m\mathbf{k}}$ given with respect to the Fermi level. The electron-phonon matrix elements contain the information about the derivative of self-consisted spin dependent electronic potential $\partial_{\mathbf{q}\nu} V^\sigma$ in the basis of Bloch functions $\psi^\sigma_{n \mathbf{k}}$:   
\begin{equation}
  g^\sigma_{mn, \nu}(\mathbf{k}, \mathbf{q}) =  \sqrt{  \frac{\hbar}{2M\omega_{\mathbf{q \nu}}}} \braket{\psi^\sigma_{m \mathbf{k+q}} \vert \partial_{\mathbf{q}\nu}V^\sigma \vert \psi^\sigma_{n \mathbf{k}}}. 
\label{eq:Coupling_g}
\end{equation}

Expectation values of $\braket{S_z}$ and $\braket{S^2_z}$ in case of $S = 1$ are the following~\cite{S_Ref3, S_Ref4, S_Ref5}: 
\begin{equation}
\begin{aligned}
  & \braket{S_z} = S \frac{1 + 2 \sum_{\mathbf{q} \nu}b_{\mathbf{q}\nu}}{ 1 + 3 \sum_{\mathbf{q} \nu}b_{\mathbf{q}\nu} + 3(\sum_{\mathbf{q} \nu}b_{\mathbf{q}\nu})^2}, \\
  & \braket{S^2_z} = \frac{2 \sum_{\mathbf{q} \nu}b_{\mathbf{q}\nu}}{ 1 + 3 \sum_{\mathbf{q} \nu}b_{\mathbf{q}\nu}} + \frac{1 + \sum_{\mathbf{q} \nu}b_{\mathbf{q}\nu}}{ 1 + 3 \sum_{\mathbf{q} \nu}b_{\mathbf{q}\nu}} \braket{S_z}
\end{aligned}
\label{eq:Magnetization}
\end{equation}

\begin{table}[!h]
\centering
\includegraphics[width=0.7\linewidth]{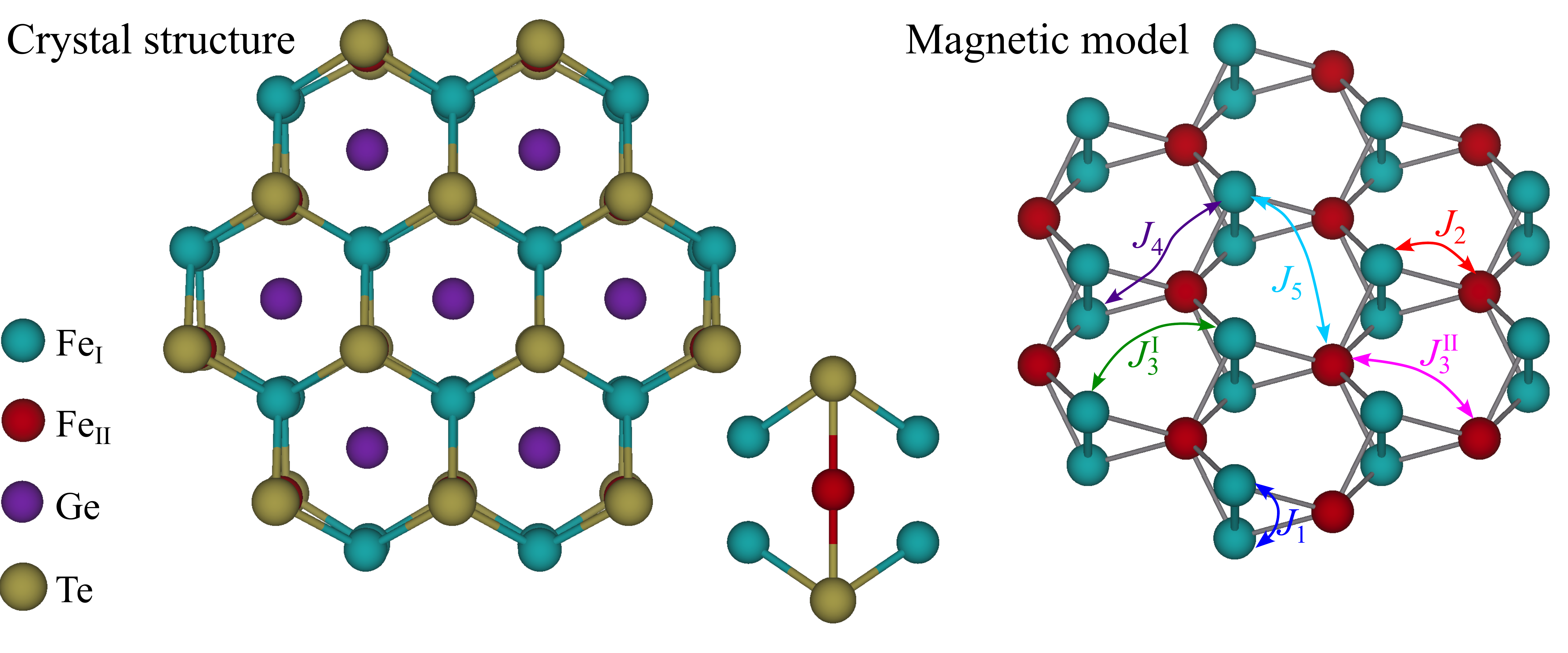}
\caption {Temperature evolution of exchange interaction (in meV) of \fgt{} due to electron-phonon coupling. }
\begin{ruledtabular}
\begin {tabular}{ccccccccc}
 $i$  &   $d(\AA)$ & $\mathrm{Fe_i - Fe_j}$  &  $J$(T = 0 K) & $J$(T = 50 K)  & $J$(T = 100 K)  & $J$(T = 200 K)  & $J$(T = 300 K) & $J(300)/J(0)$  \\
 \hline 
1 &  2.493 & I - I                &  -110.1   &  -107.4       &   -106.8    &  -105.0    &  -102.8       & 0.93        \\
2 &  2.624 & I - II               &  -37.1    &  -36.5        &   -36.2     &  -35.8     &  -35.2        & 0.95        \\
3 &  3.991 & (I - I)/(II - II)    &  1.6/7.7   &  1.6/7.3     & 1.6/7.2     & -1.6/6.9   & -1.6/6.6      & 1.0/0.85    \\
4 &  4.712 & I - I                &  1.4       & 1.2          &   1.1       & 1.0        & 0.9           & 0.64        \\
5 &  4.782 & I - II               &  -6.4      & -6.2         &  -6.1       & -5.9       & -5.6          & 0.88        \\
6 &  6.234 & I - II               &  1.6       & 1.6          &   1.6       &  1.5       & 1.4           & 0.88        \\ 
7 &  6.926 & (I - I)/(II - II)    &  -0.8/-3.6  &  -1.0/-3.4  & -0.8/-3.2   &  -0.7/-3.0 & -0.6/-2.8     & 0.75/0.77   \\
8 &  7.361 & I - I                &  -2.0       & -1.8        & -1.7        & -1.7       & -1.6          & 0.80        \\
9 &  7.998 & (I - I)/(II - II)    &  -2.0/0.5  & -2.0/0.5     & -2.0/0.4    & -1.9/0.4   & -1.9/0.3      & 0.95/0.60   \\
10& 8.377  & I - I                &  -0.8       & -0.8        & -0.8        & -0.8       & -0.7          & 0.87        \\ 
\end {tabular}
\end{ruledtabular}
\label{tab:Exchange_couplings}
\end {table}

\end{document}